\documentstyle[aps,pra,twocolumn]{revtex}
\begin{document}
\draft
%\wideabs
{
\title{Excitation-assisted inelastic processes in trapped Bose-Einstein
condensates}
\author{D. Gu\'ery-Odelin${}^1$ 
and G.V. Shlyapnikov${}^{1,2,3}$}
\address{${}^1$ Laboratoire Kastler Brossel $^{*}$,
24, Rue Lhomond, F-75231 Paris Cedex 05, France\\
${}^2$ FOM Institute for Atomic and Molecular Physics, Kruislaan 407, 
1098 SJ Amsterdam, The Netherlands \\
${}^3$ Russian Research Center, Kurchatov Institute, 
Kurchatov Square, 123182 Moscow, Russia}
\date{28 May, 1999}
\maketitle
\begin{abstract}
We find that inelastic collisional processes in Bose-Einstein condensates
induce local variations of the mean-field interparticle interaction and are
accompanied by the creation/annihilation of elementary excitations.
The physical picture is demonstrated for the case of 3-body recombination
in a trapped condensate. For a high trap barrier the production of
high-energy trapped single-particle excitations results in a strong
increase of the loss rate of atoms from the condensate.  
\end{abstract}
}
\narrowtext

Since the discovery of Bose-Einstein condensation in trapped ultra-cold 
atomic gases \cite{discovery}, inelastic collisional processes in these 
systems attract a lot of attention. Three-body recombination and two-body 
spin relaxation limit achievable densities of trapped clouds 
\cite{Amsterdam,MIT,Burt,Stamper,ENS,Cs}, and 
spin relaxation in atomic Cs even places severe limitations on achieving 
the regime of quantum degeneracy \cite{Cs}. Theoretical
studies of 3-body recombination and 2-body spin relaxation in ultra-cold 
gases provide a valuable information on the mechanisms and rates of
these processes (see \cite{J} for review, and 
\cite{Kleppner,Silvera} for the earlier work in hydrogen). These  
studies rely on a traditional approach of collisional physics in gases: 
The decay rates are found on the basis of the calculated probability of 
inelastic transition for a two- or three-body collision in vacuum.

In this Letter we find that inelastic collisional processes in 
Bose-condensed gases induce local variations of the mean-field 
interparticle interaction and can become "excitation-assisted". 
We demonstrate this phenomenon for the
case of 3-body recombination. The particles produced 
in the course of recombination have a high kinetic energy and immediately
fly away from the point of recombination. Therefore, each recombination 
event leads to an instantaneous local change of the mean field, i.e. to a 
change of the field acting on the surrounding particles. In this respect 
the recombination leads to "shaking" of the system: The time scale on which
the Hamiltonian changes is so short that the wavefunction of the system 
remains unchanged and, hence, corresponds to a superposition of many 
eigenstates of a new Hamiltonian. For this reason the 
recombination event can be accompanied by the creation or annihilation 
of elementary excitations. 

From a general point of view, the concept of "shaking" (sudden 
perturbations) was 
formulated by Migdal and used for calculating the ionization of an 
atom in $\beta$-decay (see, e.g., \cite{MK}). Sudden perturbations 
in condensed
systems are accompanied by many-body collective effects. Dilute 
Bose-Einstein condensates are unique examples of gases, as the 
mean-field interparticle interaction makes 
their behavior in many aspects similar to that of a solid.
To some extent, the picture of inelastic processes in these gases 
resembles, for example, the absorption of 
light by impurity particles in solids, where the transfer of 
impurities to an excited state
leads to a sudden local change of the polarization of the medium and,
hence, to the creation/annihilation of phonons (see, e.g., \cite{W}).

A distinct feature of "shaking" in a gaseous condensate concerns the back
action of the shaking-created excitations on the condensate.   
This effect can be strongly pronounced in a trapped condensate.
For a high trap barrier created high-energy single-particle excitations  
are still trapped and collide with condensate particles, removing them
from the condensate. Every collision thus produces already two 
energetic (trapped) atoms which again collide with the condensate etc. 
As a result, one has a cascade production of non-condensed atoms out of 
the condensate. Despite a small probability of creating 
high-energy excitations in the recombination process, this mechanism   
can significantly increase the total loss rate 
of Bose-condensed atoms. 

We will consider the most important channel of 3-body recombination in a
trapped Bose-condensed gas, that is the recombination involving 3 
condensate particles. For a single recombination event, the local change
of the interparticle interaction, $\delta H_i$, can be obtained by 
considering the center of mass ${\bf r}_i$ of the recombination-produced 
fast atom and molecule as a force center. Just before the event there are 
3 atoms at the point ${\bf r}_i$, and each particle $j$ of the sample 
interacts with this point via the potential $3U({\bf r}_j-{\bf r}_i)$. 
After the event this interaction is equal to zero, since the fast atom 
and molecule fly away from the area of recombination. 
We will use a point approximation for the interaction potential: 
$U({\bf r})\!=\!\tilde U\delta({\bf r})$, where $\tilde U\!=\!
4\pi\hbar^2a/m$, with $a$ being the $s$-wave scattering length and $m$ the 
atom mass. Then, assuming that the condensate density $n_0$ in the BEC 
spatial region greatly exceeds the density of non-condensed atoms, we 
obtain     
\begin{equation}    \label{delta}
\!\!\delta\hat H_i\!\!=\!-\!3\!\!\int\!\!d^3r_j\hat\Psi^{\dagger}\!({\bf r}_j)
U({\bf r}_j\!-\!{\bf r}_i)\hat\Psi({\bf r}_j)\!\!=\!\!-\!3\tilde U
\hat\Psi^{\dagger}\!({\bf r}_i)\hat\Psi({\bf r}_i).\!\!\!
\end{equation}
The field operator of atoms, $\hat{\Psi}\!\!=\!\!\Psi_0\!+\!\hat{\Psi}'$, 
where $\Psi_0\!=\!n_0^{\!{1/2}}$ is the 
condensate wavefunction, and $\hat{\Psi}'$ is a non-condensed part of the
operator. This part refers to quasiparticle excitations, and 
we linearize $\delta\hat H_i$ with respect to $\hat\Psi'$.
Omitting the term $-3n_0\tilde U$, which is decoupled from the quasiparticle
excitations, we obtain 
\begin{equation}     \label{delta1}
\delta\hat H_i=-3\tilde U\Psi_0( \hat{\Psi}'^\dagger({\bf r}_i)
+\hat{\Psi}'({\bf r}_i) ).
\end{equation}

The force center represents a sort of a "hole" in the coordinate space.
It has the mass $3m$ and can undergo a translational motion. Therefore,
the recombination-induced (sudden) change of the Hamiltonian $\hat H_0$
of the excitations can be written as
\begin{equation}     \label{sudden}
\delta\hat H=\int d{\bf r}_i \hat\Phi^{\dagger}({\bf r}_i)
\{\delta\hat H_i-(\hbar^2/6m)\Delta\}\hat\Phi({\bf r}_i),
\end{equation}
where $\hat\Phi({\bf r})$ is the field operator of the force center. The
first term in the rhs of Eq.(\ref{sudden}) is related to the interparticle
interaction, and the second term to the motion of the force center. 

We assume that the kinetic energy of fast particles produced 
in the recombination process greatly exceeds any other energy scale in the 
problem, and hence the creation or annihilation of excitations does not
influence the energy conservation law for the recombination. 
Then, according to the general theory of sudden perturbations \cite{MK}, 
in each
recombination event the probability of transition of the excitation 
subsystem to a new state $f$, characterized by a different set of 
quantum numbers for the excitations, is given by $w_{if}=
|\langle i|f\rangle |^2$. The symbol $\langle i|f\rangle$ stands for
the overlap integral between the wavefunction of the initial state
$i$, which is an eigenstate of the Hamiltonian $\hat H_0$, and the 
wavefunction of the state $f$ which is an eigenstate of the new  
Hamiltonian $\hat H_0+\delta\hat H$. As $\sum_f w_{if}=1$, the 
creation/annihilation of excitations does not change the total
recombination rate.
   
In Thomas-Fermi condensates the most important is the creation of 
excitations
with energies of order the chemical potential $\mu$ or larger (see below). 
These excitations are essentially quasiclassical, and their 
de Broglie wavelength
is much smaller than the spatial size of the condensate. 
Hence, the 
probability of recombination accompanied by the creation/annihilation 
of the excitations 
can be found in the local density approximation. In other words, 
as well as the recombination without production of excitations,
this process occurs locally at a given
point ${\bf r}$ characterized by local values of
the chemical potential $\mu$ and condensate density $n_0$. 
Hence, one can 
use the Bogolyubov transformation for the spatially homogeneous case   
and represent $\Psi^{\dagger}, \Psi$ in terms of the creation/annihilation
operators $\hat b_{\bf k}^{\dagger}, \hat b_{\bf k}$ of excitations 
characterized by momentum ${\bf k}$: 
\begin{equation}     \label{Bog}
\!\!\hat{\Psi}'^\dagger({\bf r}_i)\!+\!\hat{\Psi}'({\bf r}_i)\!=\!
\frac{1}{\sqrt{V}}\!\sum_{\bf k}\!
\left(\!\frac{E_k}{\epsilon_k}
\!\right)^{\!\!{1/2}}\!\!\!\!\!\!\!(\hat b_{\bf k}^\dagger\!+
\!\hat b_{-{\bf k}})\!
\exp{(-i{\bf kr}_i)}.\!\!
\end{equation}
Here $E_k=\hbar^2k^2/2m$ is the energy of a free particle,
$\epsilon_k=(E^2_k+2n_0\tilde U E_k)^{1/2}$ is the Bogolyubov energy 
of an excitation, and $V$ is the normalization volume. The
field operator of the force center can be represented in the form 
$\hat\Phi({\bf r})=(1/\sqrt{V})\sum_{\bf q}\hat a_{\bf q}\exp{(i{\bf qr})}$, 
where $\hat a_{\bf q}$ is the creation operator for the center. 
Then, using Eqs.~(\ref{delta1}) and (\ref{Bog}), Eq.(\ref{sudden}) is
transformed to 
\begin{equation}     \label{delta2}
\!\delta\hat H\!=\!-\frac{1}{\sqrt{V}}\sum_{\bf k,q}h_{\bf k}
(\hat b_{\bf k}^\dagger+\hat b_{-{\bf k}})
\hat a^{\dagger}_{{\bf q}-{\bf k}}\hat a_{\bf q}
\!+\!\sum_{{\bf q}}\frac{\hbar^2q^2}{6m}\hat a_{{\bf q}}^{\dagger}
a_{{\bf q}},
\end{equation}
where 
\begin{equation}  \label{h}
h_{\bf k}=3\tilde U n_0^{1/2}(E_k/\varepsilon_k)^{1/2}.
\end{equation}

The first term in the rhs of Eq.(\ref{delta2}), originating from the
interparticle interaction, couples the motion of the force
center with the excitation subsystem and is responsible for 
creating/annihilating excitations in the recombination process.
Considering this term as a small perturbation, we see that a
single recombination event can be accompanied by the creation/annihilation
of one excitation. Initially the momentum of the force center 
${\bf q}\!=\!0$, 
and after the creation of the excitation with momentum ${\bf k}$ 
(annihilation of the excitation with momentum $-{\bf k}$) the center 
acquires the momentum $-{\bf k}$ and the kinetic 
energy $E_k/3$. In a single recombination event, the probabilities 
of creating and annihilating the excitation with momentum ${\bf k}$
are given by 
\begin{eqnarray}   
w(N_{{\bf k}}\rightarrow N_{{\bf k}}+1)&=&\frac{1}{V}
\frac{|h_{{\bf k}}|^2(1+N_{{\bf k}})}{(\varepsilon_k+E_k/3)^2}   \nonumber  \\
w(N_{{\bf k}}\rightarrow N_{{\bf k}}-1)&=&\frac{1}{V}
\frac{|h_{{\bf k}}|^2N_{{\bf k}}}{(\varepsilon_k-E_k/3)^2},   \nonumber  
\end{eqnarray}
where $N_{{\bf k}}=[\exp{(\epsilon_k/T)}-1]^{-1}$ are the equilibrium 
occupation numbers for the excitations at a given temperature $T$.   
Then, for the rate constant of 
recombination accompanied by the creation of excitations we obtain 
\begin{equation}     \label{alpha}
\!\alpha_{ex}\!=\!\alpha\!\!\int\!\!\frac{d^3k}{(2\pi)^3}|h_{\bf k}|^2\!
\left\{\frac{1+N_k}
{(\epsilon_k\!+\!E_k/3)^2}-\frac{N_k}{(\epsilon_k\!-\!E_k/3)^2}\!\right\}\!,
\!\!
\end{equation}
with $\alpha$ being the total (event) rate constant of recombination.
The first term in the rhs of Eq.(\ref{alpha}) corresponds to spontaneous
and stimulated creation of excitations, and the second term to their 
annihilation.

For $T\ll\mu$ the annihilation and stimulated emission of excitations
can be omitted. One can put $N_k=0$, and Eq.(\ref{alpha}) yields
\begin{equation}    \label{alpha0}
\alpha_{{\rm ex}}\approx 26\alpha (n_0a^3)^{1/2}
\end{equation}
Even with a small value for the parameter $(n_0a^3)$
the large numerical factor in front of the expression (\ref{alpha0})
may imply that the creation of excitations in the course of 3-body
recombination cannot be neglected. At the highest densities 
$n_0\approx 3\times 10^{15}$ cm$^{-3}$ of the MIT sodium experiment 
\cite{MIT} Eq.(\ref{alpha0}) gives $\alpha_{{\rm ex}}/\alpha\approx 0.2$. 

With increasing temperature, the role of annihilation of the excitations
increases. However, our calculations from Eq.(\ref{alpha}) show that 
even at $T\sim\mu$ the annihilation and stimulated emission of excitations 
give a small correction to Eq.(\ref{alpha0}). Only at $T>10\mu$ the
annihilation dominates over the emission, and $\alpha_{ex}$ (\ref{alpha})
becomes negative.

The most dramatic is the influence of created excitations on the loss rate
of Bose-condensed atoms, which we will discuss for temperatures 
$T\!\alt\!\mu$ \cite{fast}. For a high trap barrier single-particle 
excitations
with energies $\epsilon_k\!\gg\!\mu$ can be still trapped. 
These atoms act as
"bullets" penetrating the condensate. In a spherical trap this
happens once per half of the oscillation period $\pi/\omega$ \cite{bullets}. 
A characteristic time which a fast atom with velocity $v_k$ spends inside 
the condensate is $\sim R/v_k$, where $R=(2\mu/m\omega^2)^{1/2}$ is the
Thomas-Fermi radius of the condensate. Hence, the rate of elastic collisions
of the fast atom with condensate atoms is 
$\sim n_0\sigma v_k(\omega R/v_k)\sim n_0\sigma c_s$, with 
$\sigma=8\pi a^2$ being the elastic cross section, and $c_s=(\mu/m)^{1/2}$ 
the sound velocity.
In each elastic collision the fast atom transfers on average a half of 
its energy to the collisional partner and removes it from the condensate. 
One has then to deal with two energetic 
atoms, and so on. This cascade process continues until the excitation 
energy becomes of order the chemical potential $\mu$. 
Accordingly, the number of lost condensate atoms will be 
$\sim\epsilon_k/\mu$,
and the characteristic time of the cascade process, $\tau\approx
2(n_0\sigma c_s)^{-1}\log{(\varepsilon_k/\mu)}$.
At realistic densities the time $\tau$ is much smaller than the 
characteristic recombination time $\tau_r\sim (\alpha n_0^2)^{-1}$.

The behavior of the excitations produced in the cascade process depends 
on the ratio $T/\mu$.
At $T\!\ll\!\mu$ their damping time strongly increases at energies well 
below $\mu$ (the decay rate is at least much
smaller than $\mu(n_0a^3)^{1/2}\!\sim\!n_0\sigma c_s$ \cite{Rev}) and is 
likely to exceed the recombination time $\tau_r$.
Therefore, these excitations 
mostly remain undamped and no longer influence the number of atoms in the
(partially destroyed) condensate. 

Thus, one has a non-equilibrium "boiling" Bose-condensed sample:
High-energy single-particle excitations, created in the 
recombination process, initiate a significant destruction of the condensate
and the formation of a non-equilibrium non-condensed cloud. 
The corresponding loss rate of condensate atoms, $\nu=\int Ln_0^3d^3r$, is
determined by the rate constant of recombination-induced production of 
excitations with energies $\varepsilon_k\!\gg\!\mu$, 
magnified by approximately
$\epsilon_k/\mu$:
\begin{equation}    \label{rate}
L=\alpha\int\frac{d^3k}{(2\pi)^3} \left| \frac{h_{\bf k}}{\epsilon_k+E_k/3}  
\right|^2\frac{\epsilon_k}{\mu}\gamma,
\end{equation}
where the numerical coefficient $\gamma\sim 1$. A precise value of $\gamma$
depends on a detailed behavior of damping rates of the excitations  
and, hence, on the trapping geometry.

The generated non-condensed cloud has energy $\sim\mu$ per
particle and occupies the volume which is of order the Thomas-Fermi
volume of the condensate. Similarly to the condensate, this
cloud decays due to 3-body recombination and, in this respect, the quantity
$\nu$ describes extra losses of (condensate) atoms from the sample.
 
The integral in Eq.(\ref{rate}) is divergent at high energies, and one should
put an upper bound $\epsilon_k=E_B$, where $E_B$ is the trap barrier.
The inequality $E_B\gg\mu$ justifies that Eq.(\ref{rate}) indeed gives the
loss rate due to the production of high-energy excitations ($\varepsilon_k
\gg\mu$). From Eq.(\ref{rate}) we obtain
\begin{equation}   \label{rate1}
L=\alpha(n_0a^3)^{1/2}\frac{81}{\sqrt{2\pi}}
\left( \frac{E_B}{\mu}\right)^{1/2}\!\!\!\gamma.
\end{equation}
As $\mu=n_{0m}\tilde{U}$, where $n_{0m}$ is the maximum condensate density,
the rate constant $L$ is independent of the number of Bose-condensed atoms.

The direct loss rate of Bose-condensed atoms due to 3-body recombination 
is $\nu_0\!=\!\!\int 3\alpha n_0^3d^3r$, as three atoms disappear immediately 
in each recombination
event. Then, using  Eq.(\ref{rate1}) and the Thomas-Fermi density profile
$n_0(r)\!=\!n_{0m}(1\!-\!r^2/R^2)$, we express the
total loss rate of Bose-condensed atoms, 
$\nu_t\!=\!\nu_0\!+\!\int \!Ln_0^3d^3r$, through $\nu_0$:
\begin{equation}      \label{total}
\!\nu_t\!=\!\nu_0\left[ 1+\frac{216}{11\sqrt{2\pi}}(n_{0m}a^3)^{1/2}
\left(\frac{E_B}{\mu}\right)^{1/2}\!\!\!\gamma\right]\!.\!
\end{equation}

The situation is the same at $T\!\sim\!\mu$, if the cascade production
of excitations with energies $\varepsilon\!\sim\!\mu$ makes the 
quasiparticle distribution strongly non-equilibrium and prevents the
damping of these excitations caused by their interaction 
with each other 
and with the thermal cloud. The number of excitations produced in one 
cascade process is $\!\sim\!E_B/\mu$, and the number of thermal 
quasiparticles with $\varepsilon\!\sim\!\mu$ is 
$N_{\mu}\!\sim\!(\mu/\hbar\omega)^3$. Thus, under the condition 
$E_B\!>\!\mu N_{\mu}$ one also
has a non-equilibrium "boiling" Bose-condensed sample, and the loss rate
of condensate atoms will be determined by Eq.(\ref{total}).

Our calculations assume the $s$-wave scattering limit $k|a|\!\ll\!1$ 
\cite{limit}, and hence
the maximum trap barrier for which they are valid is $E_B\!\approx\! 
\hbar^2/2ma^2$.
In the case of $^{87}{\rm Rb}$, the triplet scattering length is
$a\!=\!5.8$ nm, and $E_B\!=\!75$ $\mu$K. 
Assuming $\gamma\!\approx\!1$, this gives $L_t\!\approx\!3L_0$ and 
shows a qualitative significance of our
mechanism: the loss rate of Bose-condensed atoms is essentially magnified
by the creation of high-energy excitations and their destructive influence
on the condensate.
To be more quantitative, one has to consider the kinetics of excitations
produced in the sample by the initially high-energy (trapped) atom.

In ongoing BEC experiments a characteristic temperature of a Bose-condensed
sample is in the range from $100$ nK to $1$ $\mu$K and, hence, the above 
estimated
magnification of the loss rate of the condensate atoms (factor 3 for
$E_B\approx 75 \mu$K)  
practically corresponds to switching off the evaporative cooling.
With evaporative cooling on, the ratio $E_B/\mu$ for 
temperatures smaller than $\mu$ is in practice ranging from 2 to 5.
Then, at typical densities $n_0\sim 10^{14}$ cm$^{-3}$ Eq.(\ref{total}) 
only gives a 10\% increase of the total loss rate of Bose-condensed 
atoms compared to $L_0$. 
To some extent this explains the recent experiments \cite{private}, 
where a strong increase of the 3-body losses in the condensate has been 
observed after switching off the evaporative cooling.

For $T\sim\mu$ one can also think of the situation, where the cascade 
production of excitations with energies of order $\mu$ does not 
significantly destroy the equilibrium distribution of quasiparticles
in the sample. This should be the case if $E_B\ll\mu N_{\mu}$.
Then the damping of these excitations comes into play, continuously
decreasing their energy and partially refilling the condensate. 
This damping originates from (inelastic) 
scattering of a thermal excitation
on a given excitation, which transfers them to the
condensate particle and the thermal
excitation with higher energy \cite{Rev,FSW}. 
A characteristic damping rate is of order $\varepsilon (n_0a^3)^{1/2}$,
and even for the lowest excitations ($\epsilon\sim\hbar\omega$) it
can be larger than the rate of recombination. 

Consequently, one can conclude that  
the energy of excitations produced 
in the recombination process is thermalized in the gas. 
The Bose-condensed sample
will be in quasiequilibrium, with continuously increasing temperature. 
This provides extra losses of Bose-condensed atoms. 
Due to refilling the condensate in the course of
damping of the excitations, 
these losses will be smaller than
the extra losses described by Eq.(\ref{rate1}) in the  
case of a non-equilibrium "boiling" condensate.      

The rate of energy transfer from the excitations, produced in the 
recombination process, to the thermal cloud determines the increase of 
the internal energy $U$ of the gas. One can write it as 
$\dot U=\int Wn_0^3d^3r$, where the quantity $W$ is obtained in the same way 
as Eq.(\ref{rate}):
\begin{equation}     \label{W}
W=\alpha\int\frac{d^3k}{(2\pi)^3} \left| \frac{h_{\bf k}}{\epsilon_k+E_k/3}  
\right|^2\epsilon_k.
\end{equation}
Relying on Eq.(\ref{W}) and the known expressions for $U$ and the 
number of Bose-condensed atoms $N_0$ as functions of $T$ and the total
number of particles (see \cite{Rev}), we have calculated the extra 
losses of condensate atoms $|\partial N_0/\partial T|\dot{T}$, related
to the increase of temperature. At initial 
density $n_0\sim 10^{14}$ cm$^{-3}$ they do not exceed $10\%$.  

In conclusion, we have found that inelastic collisional processes in
Bose-Einstein condensates can be accompanied by the creation of elementary 
excitations. It is worth mentioning that this phenomenon is not related
to BEC as itself. It originates from the presence of the mean-field 
interparticle interaction and will also occur in a non-condensed ultra-cold gas,
as soon as the parameter $na^3$ is not extremely small.
We have revealed the influence of the production of high-energy excitations
in the course of 3-body recombination on the loss rate of atoms from a 
trapped condensate. This effect is especially pronounced for a high trap barrier
$E_B$, and
it would be valuable to perform a systematic experimental investigation 
of the loss rate of condensed atoms as a function of $E_B$.

We acknowledge fruitful discussions with J. Dalibard and M.A. Baranov.
This work was financially supported by the Stichting voor Fundamenteel 
Onderzoek der Materie (FOM), by the CNRS, by INTAS, and by the 
Russian Foundation for Basic Studies. 

$^{*}$ L.K.B. is an unit\'e de recherche de l'Ecole Normale Sup\'erieure
et de l'Universit\'e Pierre et Marie Curie, associ\'ee au CNRS.


\begin{references}
\bibitem{fast} We consider the conditions where the mean free path of
the recombination-produced fast atoms and molecules is much larger 
than the spatial size of the sample, and, being non-trapped, they 
leave the sample without collisions.
\bibitem{limit} This inequality implicitly assumes that $|a|$ is of the
order of or larger than the characteristic radius of interaction
between atoms, which is the case for alkali atom gases.
\end{references}
\end{document}